\documentclass[letterpaper,12pt]{article}
\usepackage{color,amssymb,enumerate,float,amsmath}

\usepackage{ifpdf}
\ifpdf
	\usepackage[pdftex]{graphicx}
	\usepackage[pdftex,unicode,implicit]{hyperref}

	\hypersetup{
  	pdftitle     = {}, 
  	pdfkeywords  = {},
  	pdfauthor    = {},
  	pdfcreator   = {pdf\LaTeXe\ with package \flqq hyperref\frqq},
  	pdfproducer  = {pdf\LaTeXe\ with package \flqq hyperref\frqq},
  	pdfpagemode  = UseNone,  
  	pdffitwindow = true,  
  	unicode      = true,
  	plainpages   = true,
  	colorlinks   = true,  
  	citecolor    = black,  
  	urlcolor     = blue, 
  	linkcolor    = black
	}

\else

  \usepackage[dvips]{graphicx}

\fi 

\makeatletter
\@addtoreset{equation}{section}
\makeatother

\begin{document}
	
\thispagestyle{empty}

\begin{center}
{\bf \LARGE 
	Asymptotic flatness and nonflat solutions in the critical $2+1$ Ho\v{r}ava theory
}
\vspace*{15mm}

{\large Jorge Bellor\'in, Claudio B\'orquez and Byron Droguett} 
\footnote{ 
	 Email: {\tt jorge.bellorin@uantof.cl}, \, {\tt cl.borquezg@gmail.com}, \, {\tt byron.droguett@ua.cl}}

\vspace{3ex}

{\it Department of Physics, Universidad de Antofagasta, 1240000 Antofagasta, Chile.}

\vspace{3em}

{\bf Abstract}
\begin{quotation}{\small\noindent
The Ho\v{r}ava theory in $2+1$ dimensions can be formulated at a critical point in the space of coupling constants where it has no local degrees of freedom. This suggests that this critical case could share many features with $2+1$ general relativity, in particular its large-distance effective action that is of second order in derivatives. To deepen on this relationship, we study the asymptotically flat solutions of the effective action. We take the general definition of asymptotic flatness from $2+1$ general relativity, where an asymptotically flat region with a nonfixed conical angle is approached. We show that a class of regular asymptotically flat solutions are totally flat. The class is characterized by having nonnegative energy (when the coupling constant of the Ricci scalar is positive). We present a detailed canonical analysis on the effective action showing that the dynamics of the theory forbids local degrees of freedom. Another similarity with $2+1$ general relativity is the absence of a Newtonian force. In contrast to these results, we find evidence against the similarity with $2+1$ general relativity: we find an exact nonflat solution of the same effective theory. This solution is out of the set of asymptotically flat solutions. 
}
\end{quotation}
\end{center}

\thispagestyle{empty}

\newpage
\section{Introduction}
In gravitation two features are commonly associated: the absence of propagating degrees of freedom and the triviality of the curvature, or at most, constant curvature in the presence of a cosmological constant. Of course, this is the case of 2+1 general relativity. Nontrivial configurations are more likely associated to global effects, like the case of the  Ba\~nados-Teitelboim-Zanelli black hole with negative cosmological constant \cite{Banados:1992wn}. In this paper we explore on the relationship between the absence of local degrees of freedom and the triviality of the curvature in the framework of the Ho\v{r}ava gravity \cite{Horava:2009uw,Horava:2008ih}. This theory is based on a foliation of spacelike surfaces along a given direction of time, and the symmetry is given by the transformations that preserve the foliation. A spacetime metric is not mandatory as fundamental object. Nevertheless, the gravitational fields are taken from the Arnowitt-Deser-Misner (ADM) decomposition of general relativity. The underlying gauge symmetry leads to a quantum theory with improved behavior in the ultraviolet, since terms of higher order in spatial derivatives can be incorporated in the action. Unitarity can be safe since no terms of higher order in time derivatives are neccesary to define the theory.

The most studied formulation of the Ho\v{r}ava theory, both in its projectable and nonprojectable versions (the last one extended in Ref.~\cite{Blas:2009qj}), propagates one physical degree of freedom additional to the ones of general relativity. Although this can be generically associated to the fact that the gauge symmetry group of the Ho\v{r}ava theory is smaller than the one of general relativity, actually this is not a unavoidable feature. There is a critical point in the space of coupling constants where the extra physical mode disappears (with an enhancement of the gauge symmetry the extra mode also disappears \cite{Horava:2010zj,Zhu:2011yu}). The kinetic term of the Lagrangian has the general form $\sqrt{g} N ( K_{ij} K^{ij} - \lambda  ( g^{ij} K_{ij} )^2 ) $, where $\lambda$ is an arbitrary coupling constant. The additional extra mode is eliminated when $\lambda$ takes the critical value $\lambda = 1/d$, where $d$ is the spatial dimension. To distinguish this special case, we refer to the theory formulated under the $\lambda =1/d$ condition as the critical Ho\v{r}ava theory. The canonical formulation characterizing the degrees of freedom in the critical $d=3$ case can be found in Ref.~\cite{Bellorin:2013zbp}. Two additional second-class constraints arise in the critical case, which are not associated to gauge symmetries. These constraints eliminate the extra mode. Thus, the elimination of the extra mode in the critical case is a consecuence of the dynamics, rather than the symmetry.

In a foliation of $d=2$ spatial dimensions, one expects that the critical Ho\v{r}ava theory does not propagate any mode at all, since the would-be single scalar mode should be eliminated by the additional second-class constraints. Therefore, the critical Ho\v{r}ava theory in 2 spatial dimensions behaves like $2+1$ general relativity in the sense that local degrees of freedom are absent. Since this is a central feature of the critical theory, we study this rigorously by means of a canonical analysis on the large-distance effective action, which is of second order in time and spatial derivatives. The Hamiltonian formulation is also useful to discuss the gravitational energy and its relation to the asymptotically flat configurations.  

The absence of degrees of freedom raises a question about the solutions: whether in the critical $2+1$ Ho\v{r}ava theory are or are not nonflat solutions. The qualitative comparison with general relativity is particularly relevant for the large-distance effective action, since they are of the same order. In principle, there is no reason to think  that the vacuum field equations (without cosmological constant) lead to zero curvature, since the field equations of the effective Ho\v{r}ava theory are different to the ones of general relativity, even for the critical Ho\v{r}ava theory. The ADM fields enter in some combinations that do not belong to the spacetime curvature. This can be casted on the equivalent formulation of the Einstein-aether theory \cite{Jacobson:2000xp}. If the aether field is restricted to be hypersurface orthogonal, the resulting theory is physically equivalent to the large-distance effective theory of the nonprojectable Ho\v{r}ava theory \cite{Blas:2009ck,Jacobson:2010mx,Jacobson:2013xta}. Since the gravitational field equations of the Einstein-aether theory incorporate the energy-momentum tensor of the aether field, it is clear that, whenever solutions with nonzero aether energy-momentum tensor exist, these solutions have nontrivial spacetime curvature. We remark that, in the framework of the Ho\v{r}ava theory, the aether energy-momentum tensor should not be interpreted as an external source, since it is the way of representing the intrinsically gravitational terms of the Ho\v{r}ava theory that are different to the ones of general relativity.

Asymptotically flat solutions are of special importance. They represent the gravitational field of isolated sources. In $2+1$ general relativity there is the definition of asymptotic flatness given in Ref.~\cite{Ashtekar:1993ds}, which is motivated by the rest-particle solution of Ref.~\cite{Deser:1983tn}. In contrast to the $3+1$ dimensions, in the $2+1$ case the dominant mode in the asymptotic expansion is not fixed to an unique metric at infinity. Instead, the asymptotically flat solutions approach to the asymptotic region of a cone of variable conical angle, which can be a defficit or zero angle. The case of excess angle is discarded by the condition of positive energy. The same definition of asymptotic flatness can be adopted for the $2+1$ Ho\v{r}ava theory \cite{Bellorin:2019zho}. Therefore, to undertake a general analysis on asymptotically flat configurations, one should take the general definition with the variability on the dominant mode. Moreover, in the case of the $2+1$ Ho\v{r}ava theory, there are more possibilities for positive energy among the asymptotically flat configurations. A property of the critical theory is that the coupling constant of the spatial Ricci scalar in the Lagrangian, denoted by $\beta$, has no defined sign. Due to this, the set of possible asymptotically flat solutions with positive energy is enhanced, allowing that also the metrics that approach a cone with an excess angle have positive energy.

We analyse both the degrees of freedom and the asymptotically flat solutions of the $2+1$ nonprojectable Ho\v{r}ava theory defined at the critical point $\lambda = 1/2$. We develop a detailed canonical analysis, showing the self-consistency of the theory and the fact that it has no propagating degrees of freedom. In the analysis of the solutions, we emphasize on the presence or absence of nonflat solutions. We find that a class of the asymptotically flat configurations are necessarily globally flat. This is the class of solutions that has nonnegative energy when the range $\beta > 0$ is considered in the space of coupling constants. Another related feature is the absence of a Newtonian force, as happens in $2+1$ general relativity. Despite these results, we also find that there are more possibilities for nonflat solutions: by relaxing the boundary conditions we find a nonflat solution that is not asymptotically flat. This solution is static and with rotational symmetry. It is an evidence that the critical $2+1$ Ho\v{r}ava theory is a gravitational theory without local degrees of freedom but that still possesses nonflat solutions at the level of the second-order action.

The noncritical $2+1$ Ho\v{r}ava theory has been previously studied in many aspects. This theory propagates a scalar mode, hence the dynamics is different to the case we study here. A dynamical analysis for the noncritical case was done in Ref.~\cite{Sotiriou:2011dr}, where the physical propagating mode was characterized. Several solutions has been studied in the three-dimensional noncritical Ho\v{r}ava theory. Some of these works can be found in Refs.~\cite{Park:2012ev,Shu:2014eza,Sotiriou:2014gna,Basu:2016vyz}, where black holes and other solutions with diverse asymptotics have been studied. Among them, different models of Lagrangians have been adopted, including a cosmological constant term, which allows for more kinds of asymptotic geometries, i.~e.~de Sitter, anti-de Sitter and Lifshitz asymptotics. Quantum aspects of the $2+1$ Ho\v{r}ava theory has been developed, for example, in Refs.~\cite{Benedetti:2013pya,Barvinsky:2015kil,Griffin:2017wvh,Barvinsky:2017kob,Bellorin:2019gsc}. The three-dimensional Ho\v{r}ava gravity has been related to the gauging of some nonrelativistic algebras in Ref.~\cite{Hartong:2016yrf}.

This paper is organized as follows: in section 2 we perform the canonical analysis, starting with the nonperturbative general analysis and then ending with a perturbative analysis under which the constraints can be solved explicitly. In section 3 we first study the asymptotically flat solutions, and then we find an explicit nonflat solution. We present some conclusions and in the appendix we discuss the Newtonian potential.

\section{The absence of local degrees of freedom}

\subsection{Nonperturbative canonical formulation}
We perform a detailed canonical analysis on the large-distance effective action of the critical Ho\v{r}ava theory in two spatial dimensions, with the aim of showing that this theory has no local physical degrees of freedom. A foliation of spacelike surfaces along a given direction of time is assumed. We may use local coordinates $(t,\vec{x})$ on the foliation. The theory is defined in terms of the ADM variables $N(t,\vec{x})$, $N_i(t,\vec{x})$ and $g_{ij}(t,\vec{x})$. We consider the nonprojectable case where the lapse function $N$ is allowed to depend on the time and the spatial point. We consider the effective action for large distances, which has a potential of $z=1$ order, according to the criterium of anisotropy introduced in Ref.~\cite{Horava:2009uw}. The action of the purely gravitational theory (without sources) is
\begin{equation}
S= 
\int dt d^2x \sqrt{g} N \left(G^{ijkl}K_{ij}K_{kl}
+ \beta R + \alpha a_i a^i \right)
 \,.
\label{action}
\end{equation}
In this section we use the standard notation of Riemannian geometry to denote the objects associated to the two-dimensional spatial metric $g_{ij}$. Thus, $K_{ij}$ is the extrinsic curvature of the leaves,
\begin{equation}
\label{k} K_{ij}=\frac{1}{2N}(\dot{g}_{ij}-2\nabla_{(i}N_{j)}) \,,
\end{equation}
$K$ is its trace, $K = g^{ij} K_{ij}$, and the dot stands for derivative with respect the time. The hypermatrix $G^{ijkl}$ is defined by 
\begin{equation}
\label{G} G^{ijkl}=\frac{1}{2}\left(g^{ik}g^{jl}+g^{il}g^{jk}\right)-\lambda g^{ij}g^{kl} \,.
\end{equation}
It contains the arbitrary constant $\lambda$, which is the coupling constant of the kinetic term \cite{Horava:2009uw}. $R$ is the spatial Ricci scalar. $a_i$ is the FDiff-covariant vector $a_i = \partial_i \ln N$ \cite{Blas:2009qj}. $\lambda$, $\beta$ and $\alpha$ are the independent coupling constants of the $z=1$ model.

From the identity in two spatial dimensions,
\begin{equation}
G^{ijkl}g_{kl}=(1-2\lambda)g^{ij} \,,
\label{criticalcondition}
\end{equation}
it follows that for the value  $\lambda=1/2$ the hypermatrix $G^{ijkl}$ is not invertible. This critical condition has profound consequences on the dynamics of the theory; this is our case of interest in this paper.

The condition of asymptotic flatness in this theory is the same of $2+1$ general relativity \cite{Ashtekar:1993ds}. If $x^1,x^2$ are Cartesian coordinates at spatial infinity and $r=\sqrt{x^k x^k}$, an asymptotically flat configuration behaves asymptotically as
\begin{eqnarray}
&&
g_{ij} = r^{-\mu} \left( \delta_{ij} + \mathcal{O}(r^{-1}) \right) \,,
\label{asympgij}
\\ &&
N = 1 + \mathcal{O}(r^{-1}) \,, 
\label{asympn}
\\ &&
N^i = \mathcal{O}(r^{-1}) \,,
\quad
N_i = r^{-\mu} \mathcal{O}(r^{-1}) \,,
\label{asympni}
\end{eqnarray}
where $\mu$ is an arbitrary constant that takes different values among the asymptotically flat configurations. The reason for using in the $2+1$ Ho\v{r}ava theory the same condition of asymptotic flatness as in $2+1$ general relativity is that the gravitational field of a point particle at rest is the same, and this solution is taken as the asymptotic reference to define the asymptotically flat condition. Since this is connected with the issue of the Newtonian potential, in the appendix we give more details on how this solution arises in Ho\v{r}ava theory. With the coordinate system used in (\ref{asympgij}), there are three cases for the asymptotic cone depending of the sign of $\mu$: For $\mu > 0$ a cone with a defficit angle is approached, for $\mu = 0$ the metric approaches the complete Euclidean metric without conical angle, and for $\mu < 0$ a cone with an excess angle is approached. There is an upper bound on $\mu$, $\mu < 2$ \cite{Ashtekar:1993ds}, which is needed for dynamical consistency. Below we comment that this bound is also needed in the critical Ho\v{r}ava theory.

We perform the Legendre transformation for the critical case $\lambda=1/2$ to cast the theory in its canonical formulation. The phase space is spanned by the conjugated pairs $(g_{ij},\pi^{ij})$ and $(N,P_{N})$. The action does not depend on the time derivative of lapse function $N$, hence we obtain the primary constraint $P_{N}=0$. The momentum conjugate of the spatial metric obeys the relation
\begin{equation}
\frac{\pi^{ij}}{\sqrt{g}}=G^{ijkl} K_{kl} \,.
\label{pij}
\end{equation}
Due to (\ref{criticalcondition}), the trace of this yields another primary constraint, namely,
\begin{equation}
\pi \equiv g_{ij} \pi^{ij} = 0 \,.
\label{pi}
\end{equation}
After the Legendre transformation, one integration by parts and the addition of the two primary constraints with the Lagrange multipliers $\sigma_1, \sigma_2$, we obtain the Hamiltonian
\begin{eqnarray}
H=
\int d^2x \left[ \sqrt{g} N \left( \frac{\pi^{ij}\pi_{ij}}{g}  
- \beta R - \alpha a_i a^i \right) 
+ N_{i}\mathcal{H}^{i} + \sigma_1 P_{N} + \sigma_2 \pi \right]
+ 2\pi\beta \mu \,.
\label{H}
\end{eqnarray}
The last constant term is added in order to ensure the functional differentiability of the Hamiltonian under the asymptotically flat conditions (\ref{asympgij}) -- (\ref{asympni}) \cite{Ashtekar:1993ds} (the terms that depends on $a_i$ do not affect the differentiability of the Hamiltonian, as in the noncritical $2+1$ Ho\v{r}ava theory \cite{Bellorin:2019zho}). The definition of the momentum constraint $\mathcal{H}^i$ is extended in order to get the generator of spatial diffeomorphisms on the full phase space \cite{Donnelly:2011df}, hence one defines the momentum constraint
\begin{equation}
\mathcal{H}^{i} \equiv 
-2\nabla_{k}\pi^{ik}+P_{N}\partial^{i}N = 0 \,.
\label{momentum}
\end{equation}

Next, we apply Dirac's procedure to obtain the full set of constraints. We first impose the preservation in time of the primary constraint $P_{N}=0$. This yields the Hamiltonian constraint
\begin{equation}
\mathcal{H} \equiv 
\frac{\pi^{ij}\pi_{ij}}{\sqrt{g}} - \sqrt{g}\beta R 
+ \alpha \sqrt{g} ( 2 \nabla_k a^k + \alpha a_k a^k)
= 0 \,.
\label{HamiltonianConst}
\end{equation}
The preservation of the $\pi = 0$ constraint yields a new constraint which we denote by $\mathcal{C}$. It is given by
\begin{equation}
\mathcal{C} \equiv
\frac{N}{\sqrt{g}}\pi^{ij}\pi_{ij}-\beta\sqrt{g}\nabla^{2}N 
= 0 \,.
\label{C}
\end{equation}
Dirac's procedure ends with the step of imposing the preservation of the secondary $\mathcal{H}$ and $\mathcal{C}$ constraints. These conditions generate elliptic differential equations for the Lagrange multipliers $\sigma_1$ and $\sigma_2$. The two resulting equations are equivalent to the system
\begin{eqnarray}
0 &=& 
  \beta \nabla^{2} \sigma_2
+ 4\alpha a_{k}\nabla^{k} \sigma_2
- \frac{2\alpha}{N} \nabla_{k} ( a^{k} \sigma_1 )  
+ \frac{\pi^{ij}\pi_{ij}}{g} \left( \frac{ 2\alpha }{\beta}  \frac{\sigma_1}{N} 
   - \left( 1 + \frac{2\alpha}{\beta} \right) \sigma_2 \right)
\nonumber \\ &&
+ 2 \left( 1 - \frac{\alpha}{\beta} \right) \frac{ N^2 \pi_{ij} }{ \sqrt{g} }
  a^i a^j   \,,
\\
0 &=& 
  \beta \nabla^{2}\sigma_1
- \frac{ \pi^{ij}\pi_{ij} }{g} ( \sigma_1 - N \sigma_2 )
- 2 \beta N^2  \frac{\pi_{ij}}{\sqrt{g}} 
  \left( 2 \nabla^i a^j + \left( 1 - \frac{\alpha}{\beta} \right) a^{i}a^{j} \right) 
\,.
\end{eqnarray}
In order to keep this system elliptic on $\sigma_1,\sigma_2$, we must discard the possibility of zero $\beta$, ensuring in this way the consistency of the dynamics of the theory. Therefore, the set of constraints closes consistently. In summary, the nonreduced phase space of the nonprojectable Ho\v{r}ava theory in 2 spatial dimensions and formulated at the critical point $\lambda = 1/2$ is spanned by the variables $(g_{ij},\pi^{ij})$ and $(N,P_N)$, which amount for eight functional degrees of freedom. The theory possesses the momentum constraint $\mathcal{H}^i=0$, which is a first-class constraint, and the second-class constraints $P_N = 0$, $\pi = 0$, $\mathcal{H} = 0$ and $\mathcal{C} = 0$. This gives a total of six constraints that must be imposed. Subtracting the two functional degrees of freedom corresponding to the gauge symmetry of spatial diffeomorphisms, we have that no physical, propagating, degree of freedom is left in the phase space. In other words, the reduced phase space of this theory has no dimension at all. With regards to the absence of local physical degrees of freedom, this theory behaves like 2+1 general relativity.
 
The equations of motion in the canonical formalism are
\begin{eqnarray}
\dot{N} &=& 
\sigma_1 + N^{k}\nabla_{k}N  \,,
\label{dotn}
\\
\dot{g}_{ij} &=& 
2N\frac{\pi_{ij}}{\sqrt{g}} + 2\nabla_{(i}N_{j)} + \sigma_2 g_{ij} \,,
\label{dotg}
\\
\dot{\pi}^{ij} &=& 
- \frac{N}{2\sqrt{g}}\Big(
4 \pi^{k(i}\pi^{j)}{}_{k} - g^{ij}\pi^{kl}\pi_{kl} \Big)
- \frac{\alpha}{2} \sqrt{g} N \Big(
2 a^{i}a^{j} - g^{ij} a_k a^k \Big) 
\nonumber \\ &&
+ \beta\sqrt{g} \Big( \nabla^{ij} N-g^{ij}\nabla^{2}N \Big)
- \sigma_2 \pi^{ij}
- 2\nabla_{k} N^{(i} \pi^{j)k}
+ \nabla_{k}(N^{k}\pi^{ij}) \,.
\label{dotpi}
\end{eqnarray}
To arrive at this form of the equations of motion we have considered the Hamiltonian only with the primary constraints added, as it is shown in Eq.~(\ref{H}). The secondary constraints $\mathcal{H}$ and $\mathcal{C}$ can also be added, but, with suitable boundary conditions, one finds that the solution for their corresponding Lagrange multipliers is zero, which is equivalent to drop these constraints out from the Hamiltonian (an extended discussion about this issue in the 3+1 theory can be found in \cite{Bellorin:2017gzj}). Thus, the equations of motion (\ref{dotn}) - (\ref{dotpi}) are the evolution equations of the theory on very general grounds.

Viewed as a problem of initial data, the absence of local degrees of freedom means that the constraints, together with the choice of a gauge to fix the freedom of performing spatial diffeomorphisms, determine the initial data completely. The evolution equations (\ref{dotn}) -- (\ref{dotpi}) give the flow in time fo the initial data, but the freedom to change this initial data has been already fixed by the constraints and the gauge chosen. In the next section we will see how this works explicitly, by means of a perturbative analysis.

As happens in $2+1$ general relativity \cite{Ashtekar:1993ds}, the term of the Hamiltonian (\ref{H}) that is quadratic on the canonical momentum impose an upper bound on $\mu$: in order to get a finite generator of time evolution, one must impose $\mu < 2$. Moreover, by using the constraint (\ref{HamiltonianConst}), the Hamiltonian (\ref{H}) can be written as a sum of constraints plus the boundary term,
\begin{equation}
H =
\int d^Dx 
\left( N \mathcal{H} + N_k \mathcal{H}^k + \sigma_1 P_N + \sigma_2 \pi \right) + 2 \pi \beta \mu \,.
\label{HND}
\end{equation}
The constant term $2\pi\beta\mu$, already present in (\ref{H}), is the only boundary contribution. Hence, this constant gives the value of the gravitational energy in the same way as $2+1$ general relativity. With regards to the sign of the energy and the asymptotic behavior, here there are more possibilities due to the presence of the coupling constant $\beta$, whose sign is not restricted by symmetry, physical-mode propagation nor the existence of a Newtonian potential. In general relativity the symmetry fixes $\beta = 1$. In the (linearized) $3+1$ Ho\v{r}ava theory, $\beta$ is the square of the speed of the tensorial waves, hence it must be positive. But in the critical $2+1$ theory there are no gravitational waves, since there are no local degrees of freedom. Another theory one can compare with is the $2+1$ noncritical ($\lambda \neq 1/2$) Ho\v{r}ava theory, which has a propagating mode. The squared speed of this mode is proportional to $\beta^2$, hence it does not restrict the sign of $\beta$. Another way to fix the sign of $\beta$ would be the necessity of making atractive the Newtonian potential, but in the appendix we discuss that in the critical theory there is no analogue of the Newtonian potential, as happens in $2+1$ general relativity. As a consequence, if $\beta < 0 $ one could have a configuration with $\mu < 0 $ and still it has positive energy. In $2+1$ general relativity a theorem of positivity of the energy is given in Ref.~\cite{Ashtekar:1993ds}: all globally well-defined solutions that satisfy the asymptotic condition (\ref{asympgij}) and are coupled to matter that satisfies the energy conditions have nonnegative energy. A related study for the Ho\v{r}ava theory in $3+1$ dimensions has been presented in \cite{Garfinkle:2011iw}.


\subsection{Linearized version}
It is important to have a way for solving the contraints explicitly and checking that no propagation of free data is allowed. This can be achieved conveniently in the linearized theory. In this analysis we fix $\mu = 0$ for simplicity. The absence of fluctuations on the Minkoswski background provides a clear example. The configuration that is the  analogue of the Minkowski space in 2+1 dimensions (in Cartesian coordinates) is given by the setting $N = 1$, $g_{ij} = \delta_{ij}$ and $\pi^{ij} = 0$, with multipliers $N_i = \sigma_1 = \sigma_2 = 0$. This is an exact solution of all the constraints and the evolution equations of the theory shown in the previous section. We perform perturbations around this solution by means of the canonical variables 
\begin{equation}
N=1+n \,, 
\qquad 
g_{ij}=\delta_{ij}+h_{ij} \,,
\qquad 
\pi^{ij}=p^{ij} \,.
\end{equation}
The asymptotic decay is $h_{ij}\sim\mathcal{O}(r^{-1})$, $p^{ij}\sim\mathcal{O}(r^{-2})$ and $n \sim \mathcal{O}(r^{-1})$, where $r = \sqrt{ x^i x^i }$.

We introduce the transverse-longitudinal decomposition
\begin{equation}
h_{ij}=
\left( \delta_{ij} - \frac{\partial_{i}\partial_{j}}{ \Delta } \right) h^{T} 
+ \partial_{(i}h_{j)} \,,
\end{equation}
where $\Delta = \partial_1^2 + \partial_2^2$ is the two-dimensional flat Laplacian, and similarly for $p^{ij}$. We impose  the transverse gauge $\partial_{i}h_{ij}=0$, under which the longitudinal sector of the metric is eliminated, $h_{i} = 0$. 

The linearized momentum constraint (\ref{momentum}) becomes $\partial_{i}p^{ij} = 0$, hence $p_i = 0$, whereas the linearized version of the $\pi = 0$ constraint eliminates the transverse scalar mode, $p^T=0$. Thus, in the linearized theory the canonical momentum is completely frozen. Constraint $\mathcal{C}$, given in Eq.~(\ref{C}), becomes $\beta \Delta n = 0$. Since in this perturbative analysis we are considering $n = \mathcal{O}(r^{-1})$, the solution for $n$ is $n = 0$. After this, the Hamiltonian constraint, given in Eq.~(\ref{HamiltonianConst}), becomes $\beta \Delta h^T = 0$, and since we consider $h^T = \mathcal{O}(r^{-1})$, this constraint fixes $h^T = 0$. Therefore, all the perturbations of the canonical variables are frozen by the constraints and the choice of a gauge condition, no fluctuations of the background are allowed.

Since all perturbations of the canonical variables are frozen (and the background is static), the evolution equations (\ref{dotn}) -- (\ref{dotg}) yield no relevant information, except for the fact that they fix the possible fluctuations of the Lagrange multipliers. This includes the shift vector since we have already fixed the gauge symmetry of spatial diffeomorphisms. Let us see this explicitly. We suppose that $N_i$, $\sigma_1$ and $\sigma_2$ are variables of first order in perturbations. At linear order, Eq. (\ref{dotn}) fixes $\sigma_1 = 0$, whereas Eq.~(\ref{dotpi}) yields no new information. The linearized equation (\ref{dotg}) has three different components, which are equations for the three unknowns $N_1$, $N_2$ and $\sigma_2$. Writen in matrix form, these equations are
\begin{equation}
 \left( \begin{array}{ccc}
   2 \partial_1 & 0 & 1 \\
   0 & 2 \partial_2 & 1 \\
   \partial_2 & \partial_1 & 0 
 \end{array} \right) 
 \left( \begin{array}{c}
   N_1 \\
   N_2 \\
   \sigma_2 
 \end{array} \right) = 0 \,.
\end{equation}
By multiplying this equation from the left with
\begin{equation}
 \frac{1}{2} \left( \begin{array}{ccc}
   \partial_1 & -\partial_1 & 2\partial_2 \\
   -\partial_2 & \partial_2 & 2\partial_1 \\
   2\partial_2^2 & 2\partial_1^2 & - 4 \partial_1\partial_2  
 \end{array} \right) \,,
\end{equation}
we obtain
\begin{equation}
 \Delta  \left( \begin{array}{c}
   N_1 \\
   N_2 \\
   \sigma_2 
 \end{array} \right) = 0 \,.
\end{equation}
By assuming that the Lagrange multipliers decay fast enough at infinity, we find that the only solution is $N_1 = N_2 = \sigma_2 =0$. Therefore, also the Lagrange multipliers are frozen in the linearized theory.


\section{Asymptotic conditions, flat and nonflat solutions}
We want to contrast the previously shown feature that this theory has no local degrees of freedom with the presence or absence of nonflat solutions. This can be conveniently study in the Lagrangian formulation, with the action (\ref{action}). The equations of motion derived from the action (\ref{action}) at the critical point $\lambda = 1/2$ are
\begin{eqnarray}
G^{ijkl} K_{ij} K_{kl} + \beta R + \alpha a_i a^i  
- 2 \alpha \frac{\nabla^{2} N}{N} 
&=& 0\,,
\label{deltaN}
\\ 
G^{ijkl} \nabla_{j} K_{kl}  
&=& 0 \,,
\label{deltaNi}
\\ 
\frac{1}{\sqrt{g}} \frac{\partial}{\partial t} 
\left( \sqrt{g} G^{ijkl} K_{kl} \right) 
+ 2 G^{klm(i|} \nabla_k ( N^{|j)} K_{lm} ) 
- G^{ijkl} \nabla_{n}( N^{n} K_{kl} )  
&& \nonumber 
\\
+ 2 N (K^i{}_k K^{jk} -\frac{1}{2} KK^{ij}) 
- \frac{1}{2} N g^{ij} G^{klmn} K_{kl} K_{mn} 
&& \nonumber 
\\
- \beta \left( \nabla^{ij} N - g^{ij} \nabla^{2}N \right)
+ \alpha N \left( a^i a^j - \frac{1}{2} g^{ij} a_k a^k \right)
&=& 0 \,.
\label{deltag}
\end{eqnarray}

\subsection{Asymptotically flat solutions with $\mu \geq 0$ are globally flat}

The main result about the asymptotically flat solutions can be stated as a theorem: The only regular solutions of Eqs.~(\ref{deltaN}) -- (\ref{deltag}) that satisfy the asymptotically flat condition (\ref{asympgij}) -- (\ref{asympni}) with $\mu \geq 0$ are the totally flat configurations. We remark that in the range $\beta > 0$, the configurations with $\mu \geq 0$ are the ones with nonnegative energy. 

To proof the theorem, since the spatial metric is bidimensional, we may consider a conformally flat form in Cartesian coordinates,
\begin{equation}
ds^2 = \Omega^2(t,\vec{x}) ( (dx^1)^2 + (dx^2)^2 ) \,.
\label{conformal}
\end{equation}
The leading mode of the asymptotic expansion of the spatial metric in (\ref{asympgij}) is conformal to the 2D Euclidean metric, hence we have the asymptotic behavior $\Omega^2 = r^{-\mu} \left( 1 + \mathcal{O}(r^{-1}) \right)$. An identity in the form of sum of squares arises in the $2+1$ critical theory $\lambda =1/2$ with this conformally flat form of the spatial metric: if $\mathfrak{M}_{ij}$ is an arbitrary $2\times 2$ matrix, then
\begin{equation}
 G^{ijlk} \mathfrak{M}_{ij} \mathfrak{M}_{kl} =
 \Omega^{-4} \left[ \left( \mathfrak{M}_{11} - \mathfrak{M}_{22} \right)^2
 + \left( \mathfrak{M}_{12} + \mathfrak{M}_{21} \right)^2 \right]\,.
\label{identity}
\end{equation}
We start by analysing the solution of Eq.~(\ref{deltaNi}). We recall that in the critical case $\lambda = 1/2$ we have the geometrical identity $G^{ijkl} g_{kl} = 0$. Since the spatial metric has been put in the conformally flat form (\ref{conformal}), this identity also holds for the time derivative, $G^{ijkl} \dot{g}_{kl} = 0$. Therefore, Eq.~(\ref{deltaNi}) reduces to
\begin{equation}
G^{ijkl} \nabla_j \left( N^{-1} \nabla_k N_l \right) = 0 \,.
\label{eq}
\end{equation}
We pose this equation as an equation for the shift vector $N_i$, under the condition of asymptotic flatness defined in (\ref{asympgij}) -- (\ref{asympni}). We may contract this equation with $\sqrt{g} N_i$, and integrate over the whole spatial slide at an instant of time, obtaining 
\begin{equation}
\int d^2x \sqrt{g} G^{ijkl} N_i  \nabla_j \left( N^{-1} \nabla_k N_l \right) = 0 \,.
\end{equation}
By integrating by parts we get two terms,
\begin{equation}
\int d^2x \partial_j \left( \sqrt{g} G^{ijkl} N_i  N^{-1} \nabla_k N_l \right) 
- \int d^2x \sqrt{g} N^{-1} G^{ijkl} \nabla_i N_j \nabla_k N_l = 0 \,.
\end{equation}
According to the asymptotic conditions (\ref{asympgij}) -- (\ref{asympni}), the first integral yields a boundary contribution that goes as $r^{-(2 + \mu)}$, hence it vanishes for $\mu \geq 0$. Thus, we arrive at the equation
\begin{equation}
\int d^2x \sqrt{g} N^{-1} G^{ijkl} \nabla_i N_j \nabla_k N_l = 0 \,.
\end{equation}
By using identity (\ref{identity}), this equation takes the explicit form
\begin{equation}
\int d^2x\, \Omega^{-2} N^{-1} 
\left[ \left( \nabla_1 N_1 - \nabla_2 N_2 \right)^2
+ \left( \nabla_1 N_2 + \nabla_2 N_1 \right)^2 \right] = 0 \,.  
\end{equation}
Since the integrand is a point-to-point sum of nonnegative quantities, and since we assume continuity of the functions under integration, we have that this equation necessarily implies the two equations
\begin{eqnarray}
\nabla_1 N_1 - \nabla_2 N_2 &=& 0\,,
\\
\nabla_1 N_2 + \nabla_2 N_1 &=& 0 \,.
\end{eqnarray}
Explicitly, these equations are
\begin{eqnarray}
\partial_1 N_1 - 2 \Omega^{-1} \partial_1 \Omega N_1 &=& 
\partial_2 N_2 - 2 \Omega^{-1} \partial_2 \Omega N_2 \,,
\\
\partial_1 N_2 - 2 \Omega^{-1} \partial_1 \Omega N_2 &=&
- \partial_2 N_1 + 2 \Omega^{-1} \partial_2 \Omega N_1 \,,
\end{eqnarray}
or, equivalently,
\begin{eqnarray}
\partial_1 \left( \frac{N_1}{\Omega^2} \right) &=&
\partial_2 \left( \frac{N_2}{\Omega^2} \right) \,,
\\
\partial_1 \left( \frac{N_2}{\Omega^2} \right) &=&
-\partial_2 \left( \frac{N_1}{\Omega^2} \right) \,.
\end{eqnarray}
The integrability of these equations leads to the condition of harmonic functions with respect to the totally flat Euclidean Laplacian, $\Delta \equiv \partial_1^2 + \partial_2^2$, namely 
\begin{eqnarray}
\Delta \left( \frac{N_1}{\Omega^2} \right) = 0 \,,
\quad
\Delta \left( \frac{N_2}{\Omega^2} \right) = 0 \,.
\end{eqnarray}
Note that $N_{1,2} / \Omega^2$ are $\mathcal{O}(r^{-1})$ asymptotically. The only continuous function of asymptotic order $\mathcal{O}(r^{-1})$ that is harmonic with respect to $\Delta$ is the zero function. Therefore, we have shown that the only solution of 
Eq.~(\ref{deltaNi}) satisfying the asymptotic condition (\ref{asympgij}) -- (\ref{asympni}) with $\mu \geq 0$ is $N_i = 0$. With this result the fields equations (\ref{deltaN}) and (\ref{deltag}) reduce themselves greatly. Since in this case $K_{ij} = \frac{\dot{g}_{ij}}{2N}$ and the metric has been put in conformal form, we have the identities $G^{ijkl} K_{kl} = 0$ and $K^i{}_k K^{jk} - \frac{1}{2} K K^{ij} = 0$. Equation (\ref{deltag}) takes then the form
\begin{equation}
\beta \left( \nabla^{ij} N - g^{ij} \nabla^{2}N \right)
- \frac{\alpha}{N} \left(\nabla^i N \nabla^j N 
- \frac{1}{2} g^{ij} \nabla_{k} N \nabla^{k} N \right)  
= 0 \,.
\label{asympflateq}
\end{equation}
Since the second term is traceless, the trace of this equation yields
\begin{equation}
\nabla^2 N = 0 \,.
\label{Ntrivial}
\end{equation}
By a similar procedure of multiplying this equation by $\sqrt{g} N$ and integrating over the spatial slide, we get the equation
\begin{equation}
 \int d^2x \partial_k \left(  N \partial_k N \right) 
 - \int d^2x \partial_k N \partial_k N = 0 \,.
 \label{ntrivialint}
\end{equation} 
With the asymptotic condition (\ref{asympgij}) -- (\ref{asympn}) we have that the boundary term goes as $r^{-1}$, hence it is zero. Again, the integrand of the remainning integral is nonnegative. Thus, the only continuous solution of Eq.~(\ref{ntrivialint}) is $\partial_i N = 0$, which is fixed as $N=1$ by the value at infinity. By inserting all the results we have obtained so far in the field equation (\ref{deltaN}), we obtain it reduces to the condition of zero spatial curvature, $R = 0$. After using the form of the metric (\ref{conformal}), this condition takes the explicit form 
\begin{equation}
\Omega \Delta \Omega - \partial_k \Omega \partial_k \Omega = 0 \,.
\label{eqOmega}
\end{equation}
By integrating this equation over a spatial slide and integrating by parts, we get
\begin{equation}
 \int d^2x \partial_k ( \Omega \partial_k \Omega ) 
 - 2 \int d^2x \partial_k \Omega \partial_k \Omega = 0 \,.
 \label{omegatrivial}
\end{equation}
The boundary integral is of order $\sim r^{-\mu} ( \mu + \mathcal{O}(r^{-1}) )$, hence we can ignore it. Thus, the behavior of $\Omega$ is the same as $N$: the only continuous solution of (\ref{omegatrivial}) is $\partial_i \Omega = 0$. Since $\Omega$ is constant, the only possibility left is $\mu = 0$. This completes the proof of the theorem; the only regular asymptotically flat solutions of the critical $2+1$ Ho\v{r}ava theory with $\mu \geq 0$ are the totally flat configurations. They have $\Omega = N = 1$ and $N_i = 0$. This can be stated in terms of the positivity of the energy: in the range of the space of coupling constants defined by $\beta > 0$, the only regular asymptotically flat vacuum solutions that have nonnegative energy are the flat configurations, which have zero energy.


\subsection{Nonflat solutions}
\label{rotations}
We may drop prefixed boundary conditions and look for more solutions under a specific ansatz. We start by evaluating Eqs.~(\ref{deltaN}) -- (\ref{deltag}) for the case of static configurations and imposing the condition $N_i = 0$. Equation (\ref{deltaNi}) is automatically solved. Equation (\ref{deltag}) takes exactly the form given in (\ref{asympflateq}), with its trace (\ref{Ntrivial}) yielding the condition of harmonicity on $N$. But, since in this part we do not impose boundary conditions, we continue on solving the equations without fixing $N$ yet. We put together the resulting Eqs.~(\ref{deltaN}) and (\ref{deltag}), simplified by (\ref{Ntrivial}),
\begin{eqnarray}
\beta R + \alpha a_k a^k &=& 0 \,,
\label{A1} 
\\
\beta \nabla_{ij} N - \alpha N \left( a_i a_j 
- \frac{1}{2} g_{ij} a_k a^k \right)
&=& 0 \,.
\label{B11}
\end{eqnarray}
This is the system of equations that must be solved for static configurations (the trace of Eq.~(\ref{B11}) reproduces Eq.~(\ref{Ntrivial})).

Next, we introduce the ansatz of a static spatial metric with rotational symmetry, which in polar coordinates is
	\begin{equation}
	ds^{2} = f^{-1}(r) dr^{2}+r^{2} d\theta^{2} \,,
	\end{equation} 
and we assume that the lapse function depends only on the radius, $N=N(r)$. Under this ansatz, Eq.~(\ref{A1}) and the two diagonal components of the Eq.~(\ref{B11}) take the form
	\begin{eqnarray}
	\beta \frac{f'}{r f} - \alpha \left( \frac{N'}{N} \right)^{2} &=& 0 \,,
	\label{A111}
	\\
	\beta \frac{N''}{N} + \frac{\beta}{2} \frac{f' N'}{f N} 
	- \frac{\alpha}{2} \left( \frac{N'}{N} \right)^2 &=& 0 \,,
	\label{rr}
	\\
	\frac{N'}{N} \left( \alpha \frac{N'}{N} 
	+ \frac{2 \beta}{ r } \right) &=& 0 \,.
	\label{tetateta}
\end{eqnarray}
The off-diagonal component of Eq.~(\ref{B11}) vanishes identically. We will see that Eq.~(\ref{rr}) is implied by Eqs.~(\ref{A111}) and (\ref{tetateta}), hence the system is self-consistent. 
Equation (\ref{tetateta}) has only two solutions. One is $N' = 0$, hence $N = \mbox{constant}$, which inserted in Eq.~(\ref{A111}) gives also $f = \mbox{constant}$. This is a flat configuration and Eq.~(\ref{rr}) is automatically solved by it. The other possibility is the vanishing of the second factor in (\ref{tetateta}),
\begin{equation}
  \alpha \frac{N'}{N}  + \frac{2 \beta}{ r } = 0 \,.
  \label{nonflatcase}
\end{equation}
This condition requires, by consistency, that $\alpha \neq 0$. The solution for the lapse function $N$ is obtained by the direct integration of (\ref{nonflatcase}), and then the solution for $f$ by the integration of (\ref{A111}). In the integration of $N$, a multiplicative integration constant arises; it has no physical meaning since it can be absorbed by re-scaling the time parameter of the foliation. We put this integration constant equal to $1$ for simplicity. Instead, the integration constant that arises by integrating $f$ cannot be absorbed by rescaling coordinates; we denote it by $r_0$. In this way we get the exact vacuum solution
\begin{equation}
 N(r) = r^{ -2\beta/\alpha } \,,
 \quad
 f(r) = \left( \frac{ r }{ r_0 } \right)^{ 4\beta/\alpha } \,.
 \label{nonflatsol}
\end{equation}
Equation (\ref{rr}) is solved by this configuration. Functions $N$ and $f$ are singular at the spatial infinity. The singularity could be $0$ or $\infty$ depending on the sign of $\beta/\alpha$.

We may build a spacetime curvature for this last solution if we assume that a spacetime metric is built with the ADM fields, that is, $^{(3)}g_{00} = - N^2$ and $^{(3)}g_{ij} = g_{ij}$. The point we want to highlight is that the spacetime curvature is not zero for this solution. Indeed, the only nonzero component of the Ricci tensor is $^{(3)}R_{rr}$, which, together with the Ricci scalar, are
\begin{equation}
 ^{(3)}R_{rr} = - \frac{4\beta}{\alpha} r^{-2} \,,
 \quad
 ^{(3)}R = R = 
 - \frac{4\beta}{\alpha} \frac{ r^{4\beta/\alpha-2} }{ r_0^{ 4\beta/\alpha} } \,. 
\end{equation}
There is a range in the space of parameters where the three-dimensional Ricci scalar is regular in all finite points and diverges at infinity, namely
\begin{equation}
 \frac{2\beta}{\alpha} > 1 \,.
 \label{boundalpha}
\end{equation}
We may cast this range as a condition on $\alpha$. Therefore, at least in the range (\ref{boundalpha}), this solution is not asymptotically flat. In the same range of parameters, the three-dimensional Kretschmann scalar is also regular in all finite points and diverges asymptotically,
\begin{equation}
 ^{(3)}R_{\alpha\beta\gamma\delta}\, {^{(3)}}R^{\alpha\beta\gamma\delta} =
 3 \left(\frac{4\beta}{\alpha}\right)^2
   \frac{ r^{8\beta/\alpha-4} }{ r_0^{8\beta/\alpha} } \,.
\end{equation}
The same bound (\ref{boundalpha}) on $\alpha$ has arisen in the nonprojectable Ho\v{r}ava theory in other contexts, see, for example, \cite{Blas:2009qj}.

\section*{Conclusions}

We have found that the critical $2+1$ nonprojectable Ho\v{r}ava theory has properties quite similar to $2+1$ general relativity, although some differences arise. We have studied the large-distance effective action, hence these effects manifest themselves at second order in derivatives. We have shown, by means of a rigorous canonical analysis, that this formulation of the Ho\v{r}ava theory does not propagate any physical local mode, like $2+1$ general relativity. Despite the fact that the field equations are different to the Einstein equations, we have found that, in the theory without sources, a class of regular asymptotically flat solutions are indeed globally flat. Moreover these solutions are the ones that can be found among the asymptotically flat configurations with positive energy when $\beta > 0$. Thus, one sees that, in spite of being a different theory with a different gauge symmetry group, the critical theory in two spatial dimensions tends to exhibit a dynamical behavior similar to general relativity. Another feature that we have presented is the absence of Newtonian potential (assuming again the condition of asymptotic flatness). 

It would be interesting to elucidate if the condition of asymptotic flatness leads always to totally flat solutions, by completing the proof for negative $\mu$. In the contrary case, if regular asymptotically flat solutions with negative $\mu$ exist, then they possess positive energy in the range $\beta < 0$. The sign of $\beta$ is not restricted, at least by appealing to the symmetry, wave propagation or Newtonian force. This is a particular feature of the $2+1$ Ho\v{r}ava theory. The only restriction is $\beta \neq 0$, which is a requisite for the consistency of the dynamics.

On the other hand, we have seen that the critical theory admits an exact vacuum solution with nontrivial curvature that is not asymptotically flat. We have presented the specific solution, which is static and with rotational symmetry. Thus, there is a relationship between the choice of the boundary conditions and the closedness to general relativity.

\appendix
\section{Newtonian force}
\label{sec:newton}
We may ask whether there is a place for a Newtonian force in this theory. The computations can be taken from Ref.~\cite{Bellorin:2019zho}, since the solution is the same for the critical $\lambda = 1/2$ and noncritical $\lambda \neq 1/2$ cases. Here we summarize the solution with the aim of showing explicitly the coupling between the critical Ho\v{r}ava gravity and the particle and how the Newtonian potential is lacked.

We couple the gravitational theory to a massive particle at rest. We assume that the dynamics of the particle is governed by an action of relativistic nature. Hence we assume that the action of the particle is proportional to the lenght of its worldline, embedded in a spacetime ambient. The ambient is taken from a solution of the Ho\v{r}ava theory.  We choose the time coordinate $t$ of the ambient foliation to parameterize the worldline of the particle. The mechanics of the particle is characterized by the embedding fields $q^0 = q^0(t)$ and $q^i = q^i(t)$, which define the position of the particle in the foliation. Thus, the combined system critical $2+1$ Ho\v{r}ava gravity--point particle is given by the action
\begin{equation}
S = 
\frac{1}{2\kappa} \int dt d^2x 
\sqrt{g} N \left( K_{ij}K^{ij} - \frac{1}{2} K^2 
+ \beta R + \alpha a_k a^k \right) 
- m \int dt \sqrt{ L } \,,
\label{actionparticle}
\end{equation}
where  
\begin{equation}
L = 
(N^{2} - N_{k}N^{k}) \left( \dot{q}^0 \right)^2 
- 2 N_k \dot{q}^0 \dot{q}^k - g_{kl} \dot{q}^k \dot{q}^l \,,
\label{L}
\end{equation} 
$m$ is the mass of the particle and $\kappa$ is a coupling constant. $L$ is the squared line element of the particle evaluated on the background of the ADM variables, and these variables are evaluated at the position of the particle in $L$. The field equations of the ADM variables are obtained by varying the action (\ref{actionparticle}) with respect to them. We shown the resulting equations directly evaluated on the gauge $N_i = 0$,
\begin{eqnarray}
&& 
K^{ij}K_{ij} - \frac{1}{2} K^{2} + \beta R + \alpha a_i a^i  
- 2 \alpha \frac{\nabla^{2} N}{N} 
=
2 \kappa m \frac{ (\dot{q}^0)^2 N }{ \sqrt{gL} } 
\delta^{(d)} (x^k - q^k) \,,
\label{deltaNparticle}
\\ && 
G^{ijkl} \nabla_{j} K_{kl}  
=
- \frac{ \kappa m }{ \sqrt{gL} } \dot{q}^0 \dot{q}^i \delta^{(d)} (x^m - q^m) \,,
\label{deltaNiparticle}
\\ 
&& 
\frac{1}{\sqrt{g}} \frac{\partial}{\partial t} 
\left( \sqrt{g} G^{ijkl} K_{kl} \right) 
+ 2 N (K^i{}_k K^{jk} - \frac{1}{2} KK^{ij}) 
- \frac{1}{2} N g^{ij} G^{klmn} K_{kl} K_{mn} 
\nonumber 
\\ &&
- \beta \left( \nabla^{ij} N - g^{ij} \nabla^{2}N \right)
+ \frac{\alpha}{N} \left(\nabla^i N \nabla^j N 
- \frac{1}{2} g^{ij} \nabla_{k} N \nabla^{k} N \right)  
\nonumber 
\\ &&
=
\frac{ \kappa m}{ \sqrt{gL}} \dot{q}^{i} \dot{q}^{j} \delta^{(d)} (x^m - q^m) \,.
\label{deltagparticle}
\end{eqnarray}
In the $N_i = 0$ gauge we have that $K_{ij} = \dot{g}_{ij}/2N$. The equations of motion corresponding to the variations of the coordinates of the particle are
\begin{eqnarray}
{q}^i{}'' + \Gamma^{i}_{kl} {q}^k{}' {q}^l{}' 
+ \frac{1}{2} \partial^i N^2 
( {q}^0{}' )^2
&=& 0 \,,
\label{partspace}
\\
\frac{2}{\sqrt{L}} \frac{d}{d t} N^{2} {q}^{0}{}' 
+ \partial_0 N^2  ({q}^0{}')^2
- \partial_{0} g_{ij} {q}^{i}{}' {q}^{j}{}'	
&=& 0 \,,
\label{parttime}
\end{eqnarray}
where the prime means
\begin{equation}
{\psi}' \equiv \frac{1}{\sqrt{L}}\frac{\partial \psi}{\partial t} \,. 
\end{equation}

In the equations of motion (\ref{deltaNparticle}) -- (\ref{parttime}) we consider that the particle and its gravitational field (the $N$ and $g_{ij}$ fields) are static. A suitable choice for the location of the particle is $q^0 = t$, $q^i = 0$, which means that the particle remains at the origin.  Under these settings, Eq.~(\ref{deltagparticle}) acquires exactly the same form displayed in (\ref{asympflateq}) for the vacuum theory. Its trace gives again the condition of harmonicity on $N$ (\ref{Ntrivial}). Imposing the asymptotic condition (\ref{asympgij}) -- (\ref{asympn}), which is appropiated for the Newtonian case, we have again that the only solution is $N = 1$ everywhere. Since a nonconstant $-N^2$ would be the analogue of the Newtonian potential, we have that there is no Newtonian force in this theory. Hence, the situation is the same as in $2+1$ general relativity. Indeed, the rest of the analysis for solving the remaining field equations is parallel to general relativity. Equation (\ref{deltaNiparticle}) is automatically solved. Particle's equations of motion, given in Eqs.~(\ref{partspace}) -- (\ref{parttime}), are automatically solved considering that $N=1$. The remaining is the analogue of the time-time component of the Einstein equations, Eq.~(\ref{deltaNparticle}). It takes the form
\begin{equation}
\sqrt{g} R = \frac{2 \kappa m }{\beta} \delta^{(2)}(x^i) \,.
\label{riccidelta}	 
\end{equation}
This equation was solved in \cite{Deser:1983tn}. Its solution, in polar coordinates, is
\begin{equation}
ds^{2} = 
r^{-\frac{ \kappa m}{\pi\beta}} ( dr^2 + r^2 d\theta^2 ) \,. 
\label{polar}
\end{equation}
The generic geometry is a flat cone (other geometries are possible in the space of parameters \cite{Deser:1983tn,Ashtekar:1993ds}).

\section*{Acknowledgements}

B.~D.~and C.~B.~are partially supported by the grants CONICYT PFCHA/DOCTO-RADO BECAS CHILE/2019 - 21190398 and 21190960 respectively. B.~D.~ is partially supported by the Universidad de Antofagasta grant PROYECTO ANT1756, Chile.



\begin{thebibliography}{99}
	
  \bibitem{Banados:1992wn}
  M.~Ba\~nados, C.~Teitelboim and J.~Zanelli,
  ``The Black hole in three-dimensional space-time'',
  Phys.\ Rev.\ Lett.\ {\bf 69} 1849 (1992)
  [arXiv:hep-th/9204099 [hep-th]].
  
  \bibitem{Horava:2009uw} 
  P.~Ho\v{r}ava,
  ``Quantum Gravity at a Lifshitz Point,''
  Phys.\ Rev.\ D {\bf 79}, 084008 (2009)
  [arXiv:0901.3775 [hep-th]].
  
  \bibitem{Horava:2008ih}
  P.~Ho\v{r}ava,
  ``Membranes at Quantum Criticality'',
  JHEP \textbf{03} 020 (2009)
  [arXiv:0812.4287 [hep-th]].
  
  \bibitem{Blas:2009qj} 
  D.~Blas, O.~Pujolas and S.~Sibiryakov,
  ``Consistent Extension of Ho\v{r}ava Gravity,''
  Phys.\ Rev.\ Lett.\  {\bf 104} 181302 (2010)
  [arXiv:0909.3525 [hep-th]].  
  
  \bibitem{Horava:2010zj}
  P.~Ho\v{r}ava and C.~M.~Melby-Thompson,
  ``General Covariance in Quantum Gravity at a Lifshitz Point",
  Phys.\ Rev.\ D {\bf 82} 064027 (2010)
  [arXiv:1007.2410 [hep-th]].
  
  \bibitem{Zhu:2011yu}
  T.~Zhu, F.~W.~Shu, Q.~Wu and A.~Wang,
  ``General covariant Horava-Lifshitz gravity without projectability condition and its applications to cosmology'',
  Phys. Rev. D \textbf{85} 044053 (2012)
  [arXiv:1110.5106 [hep-th]].
  
  \bibitem{Bellorin:2013zbp}
  J.~Bellor\'in, A.~Restuccia and A.~Sotomayor,
  ``Consistent Ho\v{r}ava gravity without extra modes and equivalent to general relativity at the linearized level'',
  Phys. Rev. D \textbf{87} 084020 (2013)
  [arXiv:1302.1357 [hep-th]].
  
  \bibitem{Jacobson:2000xp}
  T.~Jacobson and D.~Mattingly,
  ``Gravity with a dynamical preferred frame''
  Phys. Rev. D \textbf{64} 024028 (2001)
  [arXiv:gr-qc/0007031 [gr-qc]].
  
  \bibitem{Blas:2009ck}
  D.~Blas, O.~Pujolas and S.~Sibiryakov,
  ``Comment on `Strong coupling in extended Ho\v{r}ava-Lifshitz gravity' '',
  Phys. Lett. B \textbf{688} 350 (2010)
  [arXiv:0912.0550 [hep-th]].
  
  \bibitem{Jacobson:2010mx} 
  T.~Jacobson,
  ``Extended Ho\v{r}ava gravity and Einstein-aether theory,''
  Phys.\ Rev.\ D {\bf 81} 101502 (2010)
  Erratum: [Phys.\ Rev.\ D {\bf 82} 129901 (2010)]
  [arXiv:1001.4823 [hep-th]].
  
  \bibitem{Jacobson:2013xta}
  T.~Jacobson,
  ``Undoing the twist: The Ho\v{r}ava limit of Einstein-aether theory,''
  Phys.\ Rev.\ D \textbf{89} 081501 (2014)
  [arXiv:1310.5115 [gr-qc]].
  
  \bibitem{Ashtekar:1993ds} 
  A.~Ashtekar and M.~Varadarajan,
  ``A Striking property of the gravitational Hamiltonian'',
  Phys.\ Rev.\ D {\bf 50} 4944 (1994)
  [arXiv:gr-qc/9406040 [gr-qc]].
  
  \bibitem{Deser:1983tn} 
  S.~Deser, R.~Jackiw and G.~'t Hooft,
  ``Three-Dimensional Einstein Gravity: Dynamics of Flat Space'',
  Annals Phys.\ {\bf 152} 220 (1984).
  
  \bibitem{Bellorin:2019zho}
  J.~Bellor\'in and B.~Droguett,
  ``Point-particle solution and the asymptotic flatness in 2+1D Ho\v{r}ava gravity,''
  Phys.\ Rev.\ D {\bf 100} 064021 (2019)
  [arXiv:1905.02836 [gr-qc]].
  
  \bibitem{Sotiriou:2011dr} 
  T.~P.~Sotiriou, M.~Visser and S.~Weinfurtner,
  ``Lower-dimensional Ho\v{r}ava-Lifshitz gravity'',
  Phys.\ Rev.\ D {\bf 83} 124021 (2011)
  [arXiv:1103.3013 [hep-th]].
  
  \bibitem{Park:2012ev}
  M.~I.~Park,
  ``The Rotating Black Hole in Renormalizable Quantum Gravity: The Three-Dimensional Ho\v{r}ava Gravity Case,''
  Phys. Lett. B \textbf{718} 1137 (2013)
  [arXiv:1207.4073 [hep-th]].
  
  \bibitem{Shu:2014eza}
  F.~W.~Shu, K.~Lin, A.~Wang and Q.~Wu,
  ``Lifshitz spacetimes, solitons, and generalized BTZ black holes in quantum gravity at a Lifshitz point,''
  JHEP \textbf{04} 056 (2014)
  [arXiv:1403.0946 [hep-th]].
  
  \bibitem{Sotiriou:2014gna}
  T.~P.~Sotiriou, I.~Vega and D.~Vernieri,
  ``Rotating black holes in three-dimensional Ho\v{r}ava gravity'',
  Phys. Rev. D \textbf{90} 044046 (2014)
  [arXiv:1405.3715 [gr-qc]].
  
  \bibitem{Basu:2016vyz}
  S.~Basu, J.~Bhattacharyya, D.~Mattingly and M.~Roberson,
  ``Asymptotically Lifshitz spacetimes with universal horizons in $(1 + 2)$ dimensions'',
  Phys. Rev. D \textbf{93} 064072 (2016)
  [arXiv:1601.03274 [hep-th]].
  
  \bibitem{Benedetti:2013pya}
  D.~Benedetti and F.~Guarnieri,
  ``One-loop renormalization in a toy model of Ho\v{r}ava-Lifshitz gravity'',
  JHEP \textbf{03} 078 (2014)
  [arXiv:1311.6253 [hep-th]].
  
  \bibitem{Barvinsky:2015kil}
  A.~O.~Barvinsky, D.~Blas, M.~Herrero-Valea, S.~M.~Sibiryakov and C.~F.~Steinwachs,
  ``Renormalization of Hořava gravity'',
  Phys. Rev. D \textbf{93} 064022 (2016)
  [arXiv:1512.02250 [hep-th]].
  
  \bibitem{Griffin:2017wvh} 
  T.~Griffin, K.~T.~Grosvenor, C.~M.~Melby-Thompson and Z.~Yan,
  ``Quantization of Ho\v{r}ava gravity in 2+1 dimensions'',
  JHEP {\bf 1706} 004 (2017)
  [arXiv:1701.08173 [hep-th]]. 
  
  \bibitem{Barvinsky:2017kob} 
  A.~O.~Barvinsky, D.~Blas, M.~Herrero-Valea, S.~M.~Sibiryakov and C.~F.~Steinwachs,
  ``Ho\v{r}ava Gravity is Asymptotically Free in $2+1$ Dimensions'',
  Phys.\ Rev.\ Lett.\  {\bf 119} 211301 (2017)
  [arXiv:1706.06809 [hep-th]].
  
  \bibitem{Bellorin:2019gsc}
  J.~Bellor\'in and B.~Droguett,
  ``Quantization of the nonprojectable 2+1D Ho\v{r}ava theory: The second-class constraints'',
  Phys. Rev. D \textbf{101} 084061 (2020)
  [arXiv:1912.06749 [hep-th]].
  
  \bibitem{Hartong:2016yrf}
  J.~Hartong, Y.~Lei and N.~A.~Obers,
  ``Nonrelativistic Chern-Simons theories and three-dimensional Ho\v{r}ava-Lifshitz gravity'',
  Phys. Rev. D \textbf{94} 065027 (2016)
  [arXiv:1604.08054 [hep-th]].
 
  \bibitem{Donnelly:2011df}
  W.~Donnelly and T.~Jacobson,
  ``Hamiltonian structure of Ho\v{r}ava gravity'',
  Phys. Rev. D \textbf{84} 104019 (2011)
  [arXiv:1106.2131 [hep-th]].
  
  \bibitem{Bellorin:2017gzj}
  J.~Bellor\'in and A.~Restuccia,
  ``On the space of solutions of the Ho\v{r}ava theory at the kinetic-conformal point'',
  Gen. Rel. Grav. \textbf{49} 132 (2017)
  [arXiv:1705.10161 [hep-th]].
  
  \bibitem{Garfinkle:2011iw}
  D.~Garfinkle and T.~Jacobson,
  ``A positive energy theorem for Einstein-aether and Ho\v{r}ava gravity'',
  Phys. Rev. Lett. \textbf{107} 191102 (2011)
  [arXiv:1108.1835 [gr-qc]].
  
  
  
 
  

  
 
  
     
\end{thebibliography}
\end{document}